\documentclass[aps,prl,twocolumn,floats]{revtex4-2}
\usepackage[margin=0.8in,footskip=0.25in,paperwidth=8in,paperheight=13.5in]{geometry}
\usepackage{latexsym}
\usepackage{amsmath}

\usepackage{amssymb}
\usepackage{bm}
\usepackage{wasysym}
\usepackage[dvips]{color}
\usepackage{graphicx}
\usepackage{subfigure}
\usepackage{epsfig}
\usepackage{balance}
\usepackage{subfigure}
\usepackage{ulem}

\DeclareMathAlphabet{\mathpzc}{OT1}{pzc}{m}{it}

\newcommand{\hide}[1]{}
\newcommand{\veps}{\varepsilon}

\def\ra{\rangle}
\def\la{\langle}

\def\veps{\varepsilon}

\usepackage{tikz}
\usetikzlibrary{matrix}
\usetikzlibrary{arrows}

\graphicspath{{./figures/}}
\usepackage{float}

\usepackage[colorlinks,bookmarks=true,citecolor=blue,linkcolor=blue,urlcolor=blue, breaklinks=true]{hyperref}
\usepackage[sort&compress]{natbib}

\begin{document}

\title{Chiral Bloch states in single layer graphene with Rashba spin-orbit coupling: Spectrum and spin current density}
\author{Y. Avishai$^{1}$ and Y. B. Band$^{2}$}
\affiliation{$^1$Department of Physics,
  Ben-Gurion University of the Negev,
  Beer-Sheva, Israel, \\
  New York University and the NYU-ECNU Institute
  of Physics at NYU Shanghai, 3663 Zhongshan Road North,
  Shanghai, 200062, China, \\
  and Yukawa Institute for Theoretical Physics, Kyoto, Japan\\
  $^2$Department of Physics, Department of Chemistry,
  Department of Electro-Optics, and
  The Ilse Katz Center for Nano-Science,
  Ben-Gurion University of the Negev,
}


\begin{abstract} 
We study the Bloch spectrum and spin physics of 2D massless Dirac electrons in single layer graphene subject to a one dimensional periodic Kronig-Penney potential and Rashba spin-orbit coupling.  The Klein paradox exposes  novel features in the band dispersion and in graphene spintronics. In particular  it is shown that: (1) The Bloch energy dispersion $\veps(p)$  has unusual structure: There are {\it two Dirac points} at Bloch momenta $\pm p \ne 0$ and a narrow band emerges between the wide valence and conduction bands. (2) The charge current and the spin density vector vanish. (3) Yet, all the non-diagonal elements of the spin current density tensor are finite and their magnitude increases linearly with the spin-orbit strength. In particular, there is a spin density current whose polarization is perpendicular to the graphene plane. (4) The spin density currents are space-dependent, hence their continuity equation includes a finite spin torque density.  
\end{abstract}

\maketitle

{\it Introduction}: Following the discovery of graphene \cite{Geim}, novel phenomena were  predicted in its electronic properties \cite{Guinea,Sarma}. Among these, the Klein paradox \cite{Klein} and chiral tunneling in single layer graphene (SLG) were reported in a seminal paper \cite{Katsnelson_06}, and further  findings were reported in Refs.~\cite{Katsnelson-2012, AF_11}.  Due to chirality near a Dirac point, electrons execute unimpeded transmission through a potential barrier even for energies below the barrier. This scenario is related to the absence of back-scattering for electron-impurity scattering in carbon nanotubes \cite{Ando}.  Several extensions were reported in Refs.~\cite{Peeters, Barbier,AB}.  In parallel, investigation of the role of electron spin in graphene led to the emergence of a new field: graphene spintronics \cite{Huertas, Min,Yao,Castro, Trau,Tombros, Tombros1,Cho, Zarea, Rashba1,Liu, Gmitra0,  Berc,Sergej1,Schwierz,Tse,Sergej2, Miao, Jo, Liu2, Mar, Richter, Patra, Dulbak, Zomer, Zhang1,Lenz, Shakouri, Tang, Jaros, Kochan, Tuan, Gmitra, Ferrari, Roch, Zhang, Drogler, Avsar, Ingla, Medina, Li, Lin, Amir, Shifei, AB2}. The role of Klein paradox in graphene spintronics
 is reported in Refs. \cite{Richter, Shakouri,AB2}, who studied electron transmission through a barrier in the presence of Rashba spin orbit coupling (RSOC).  
 
 In this work we expose yet another facet of the Klein paradox in graphene spintronics by elucidating the physics of electrons in SLG subject to a periodic one dimensional Kronig-Penney potential (1DKPP) and {\it uniform} RSOC.   Thereby the roles of the Klein paradox \cite{Katsnelson_06} and RSOC in SLG are combined with the Bloch theorem, and novel aspects of band structure and spin related observables are exposed.  Recall that RSOC can be controlled by an externally applied uniform electric field ${\bf E}=E_0 \hat{\bf z}$ perpendicular to the SLG lying in the $x$-$y$ plane, as in the Rashba model for the two-dimensional electron gas \cite{Rashba}. We hope this study will motivate further study of graphene based spintronic devices that do not rely on the use of an external magnetic field or magnetic materials. 

Observables that are calculated include the Bloch spectrum $\veps(p)$ ($p= $ crystal momentum), spin density, and spin current density (related to  spin torques \cite{Niu}).  Their properties are remarkably different from those predicted in bulk SLG in the absence of a 1DKPP, wherein the Klein paradox does not play a role): 
(1) The spin-orbit (SO) splitting of levels in the Bloch energy dispersion is rather unusual: Recall that for $\lambda=0$, there are two degenerate levels in the valence and the conduction band and the gap is closed at a single Dirac point at Bloch momentum $p=0$ [see Fig.~{\ref{Fig1}(a) below].  For $\lambda>0$ this single Dirac point is split into two points located at $\pm p \ne 0$ [see Fig.~\ref{Fig1}(c) below].
(2) Although the charge current and the spin density vector vanish, the non-diagonal elements of the spin current density tensor $J_{ij}$ are finite (here $i=x,y,z$ is the polarization direction and $j=x,y$ is the propagation direction). Thus, unlike in bulk SLG \cite{Zhang}, $J_{zx} \ne 0$ and $J_{zy} \ne 0$ (current is polarized perpendicular to the SLG plane).   
(3) $J_{ij}(x)$ is space-dependent so that there is a finite spin torque \cite{Niu}. 
(4) The response of the spin current densities to the RSOC strength $\lambda$ is substantial even for small $\lambda$ (the magnitude of $\lambda$ due to a strong perpendicular electric field in SLG as reported in Ref.~\cite{Gmitra} is a fraction of meV). 
These predictions regarding graphene spintronics are experimentally verifiable.

{\it Formalism}:
Consider a system of massless 2D Dirac electrons in SLG lying in the $x$-$y$ plane subject to a uniform electric field ${\bf E}=E_0 \hat{\bf z}$ and a 1D periodic Kronig-Penney potential, 
\begin{equation} \label{u}
 u(x)=u_0 \sum_{m=-\infty}^\infty \Theta(x-m \ell)\Theta(m \ell+d-x). 
\end{equation}  
 The (Fermi) energy $\veps$ and the potential height $u_0$ satisfy the inequality $u_0>\veps>0$ (the condition for the emergence of the Klein paradox wherein electrons propagate under the barriers). Our goal is to derive the Bloch spectrum and Bloch functions in oder to predict spin related observables. The problem is treated here within the continuum formulation near one of the Dirac points, say ${\bf K}'$. Since the transverse wave number $k_y$ is conserved, the wave function can be factored: $\Psi(x,y)=e^{i k_y y}\psi(x)$.  Recall that, in addition to the isospin ${\bm \tau}$ encoding the two-lattice structure of SLG, there is now a {\it real spin}, ${\bm \sigma}$. Hence, the wave function $\psi(x)$ is a four component spinor in ${\bm \sigma}\otimes{\bm \tau}$ (spin$\otimes$isospin) space. It has dimensions of $1/\sqrt{A}$ where $A$ is some relevant area.  Hereafter we take $A=(d+ \ell)\times 1$ (nm)$^2$, and omit this factor when no confusion arises.   The Hamiltonian is,
\begin{eqnarray} \label{1}
&& h(-i \partial_x,k_y,\lambda) = \gamma \{[-i \partial_x + \lambda (\hat {\bf z} \times {\bm \sigma})_x] \tau_x \nonumber \\
&& \ \ + [k_y+\lambda (\hat {\bf z} \times {\bm \sigma})_y] \tau_y \}+u(x) \nonumber \\ 
&&  \equiv h_0(-i \partial_x,k_y,\lambda)+u(x).   
\end{eqnarray}
which is a 4$\times$4 matrix first-order differential operator.  Here $\gamma$ = $\hbar v_F = 659.107$ meV$\cdot$nm is the kinetic energy parameter, and $\lambda$ is the RSOC strength parameter \cite{Shnirman} (it is also the inverse SO length parameter $\lambda =1/\ell_{so} \propto E_0$). The products, $\sigma_x \tau_x$, $\sigma_y \tau_y$, implicitly incorporate a Kronecker product.  $\psi(x)$ is a combination of four component plane-wave spinors, 
  $e^{\pm i k_x x}v(\pm k_x)$  \ 
 (between barriers),  \ and 
 $e^{\pm i q_x x}w(\pm q_x)$ \  (in the barriers).
The constant vectors $v(\pm k_x)$ and $w(\pm q_x)$  satisfy the algebraic linear equations,
\begin{eqnarray} \label{3}
 && h_0(\pm k_x,k_y,\lambda)v(\pm k_x)=\veps v(\pm k_x), \nonumber \\
&& h(\pm q_x,k_y,\lambda)w(\pm k_x)=\veps w(\pm k_x).  
\end{eqnarray} 
The vectors $v(\pm k_x)$ and $w(\pm k_x)$ cannot be chosen as spin eigenfunctions {\it because spin is not conserved}. Moreover, Eqs.~(\ref{3}) are not eigenvalue equations. Indeed, assuming fixed transverse wave number $k_y$, potential parameters $u_0,d,\ell$ and RSOC strength $\lambda$, the wave numbers $k_x$ and $q_x$ must depend on the (yet unknown) energy $\veps$.   For $\veps>0$ (recall the condition of the Klein paradox), and for each sign $s=\pm$, there are two wave numbers  
that solve these implicit equations: $s k_{xn}(\veps)$ for 
 $x \notin [0,d]$, and $s q_{xn}(\veps)$ for $x \in [0,d]$ ($n=1,2$).  (The ubiquitous energy dependence will be occasionally omitted). 
 Therefore, equations~(\ref{3}) are implicit equations for $ s k_{xn}(\veps)$ and $s q_{xn}(\veps)$ as well as for $v_{ns} \equiv v[s k_{xn}(\veps)]$ and 
$w_{ns} \equiv w[s k_{xn}(\veps)]$ (where $n=1,2$ and $s=\pm$). 
The solution of Eqs.~(\ref{3}) is given by,
\begin{eqnarray} \label{3kn}
 && k_{xn}^2=[\veps+(-1)^{n+1} \lambda]^2-\lambda^2-k_y^2, \nonumber \\
 && q_{xn}^2=[\veps+(-1)^{n+1} \lambda-u_0]^2-\lambda^2-k_y^2,
\end{eqnarray}
together with the vectors $v_{ns}$ and $w_{ns}$ (their analytic expressions will not be explicitly given here). They are normalized as
  $\la v_{ns}\vert v_{n's} \ra = \la w_{ns}\vert w_{n's} \ra=\delta_{nn'}$, but $\la v_{n+}\vert v_{n-}\ra \ne 0$ and $\la w_{n+}\vert w_{n-}\ra \ne 0$.

The general form of the wave functions between and within the barriers is then:
\begin{eqnarray} \label{SO3}
\psi(x) = \sum_{n, s=\pm} \begin{cases} a_{n s} e^{i s k_{xn} x} v_{n s}, \ (u(x)\!=0), \\
 b_{n s} e^{i s q_{xn} x} w_{n s}, \ (u(x)\!=u_0). \end{cases}
\end{eqnarray}
The constants $a_{ns}(\veps)$ and $b_{ns}(\veps)$ with $n=1,2$, and $s=\pm$, are determined by matching the wave functions on the walls of the barrier and employing Bloch condition to which we now turn. 
 
Consider the unit cell $[0,R]$ consisting of the barrier region $[0,d]$ and the spacing $[d,d+\ell=R]$, corresponding to the case $m=0$ in Eq.~(\ref{u}). The matching equations at the left wall of the barrier $x=0$ implies $\psi(0^-)=\psi(0^+)$. It is written in terms of $\{ a_{ns}\}, \{ b_{ns}\}$ using the following notation: 
\begin{eqnarray} \label{compact-ab} 
{\bf a}&=&(a_{1^+},a_{2^+},a_{1^-},a_{2^-})^T, \nonumber \\ 
{\bf b}&=&(b_{1^+},b_{2^+},b_{1^-},b_{2^-})^T.
\end{eqnarray}
${\bf a}$ and ${\bf b}$ are the $4$$\times$$1$ column vectors of coefficients introduced in Eq.~(\ref{SO3}). Moreover, 
\begin{eqnarray} \label{com pact-UV}
V &=& (v_{1^+},v_{2^+},v_{1^-},v_{2^-}), \nonumber \\
W &=& (w_{1^+},w_{2^+},w_{1^-},w_{2^-}),
\end{eqnarray}
are $4$$\times$$4$ matrices built from the $4$$\times$$1$ column vectors introduced in Eq.~(\ref{SO3}). The matching equations at $x=0$ and the transfer matrix carrying $\psi(0^-)$ to $\psi(0^+)$ are then given by,
\begin{equation} \label{SO7}
  V{\bf a} = W{\bf b}, \ \Rightarrow \ T_{0^- \to  0^+} = W^{-1}V, 
\end{equation} 
so that $T_{0^- \to  0^+}{\bf a}={\bf b}$. Similarly, the transfer matrix carrying $\psi(d^-) \to \psi(d^+)$ across the right wall of the barrier is $T_{d^- \to  d^+}=V^{-1}W$. To complete the construction of the transfer matrix $T$ that carries the wave function across a unit cell from $x=0^-$ to $x=R^-= \ell+d^-$ recall that the propagation of $\psi(x)$ from $0^+ \to d^-$ and from $d^+ \to R^-$ is respectively controlled by the 4$\times$4 diagonal phase-factor matrices, 
\begin{eqnarray} 
\Phi_q&=&\mbox{diag}[e^{i q_{x1} d},e^{i q_{x2} d},e^{-i q_{x1} d},e^{-i q_{x2} d}], \nonumber \\
\Phi_k&=&\mbox{diag}[e^{i k_{x1} \ell},e^{i k_{x2} \ell},e^{-i k_{x1} \ell},e^{-i k_{x2} \ell}],
\label{SO8}
\end{eqnarray} 
which leads eventually to the expression,
$T=\Phi_kT_{d^-\to d^+}\Phi_qT_{0^- \to 0^+}$.
$T$ is a symplectic $4$$\times$$4$ matrix satisfying
$\mbox{Det} [T]=1$ and $T^\dagger \Sigma_z T=\Sigma_z$, where $\Sigma_z={\bf 1}_{2\times 2}\otimes \tau_z$.  The Bloch theorem (for fixed $\lambda, k_y,u_0,d,\ell$) requires that $\psi(x+R) = e^{i p R} \psi(x)$ where $p$ is the crystal wave number. This implies the eigenvalue equation
\begin{equation}  \label{SO11}
  T(\veps) {\bf a}(\veps)=e^{i p R}{\bf a(\veps)}.
\end{equation}
Equation (\ref{SO11}) defines a relation between the four eigenvalues 
$\{ \lambda_j(\veps) \}$ ($j=1,2,3,4$) of $T(\veps)$  and the Bloch wave number $p$, that is, Im$[\lambda_j(\veps)]=\sin pR$.   Thereby we get the dispersion curves $\veps_j(p)=[\lambda^I_j]^{-1} \big (\sin pR \big )$.  The eigenvalues of $T$ satisfy the equalities $\lambda_1 = 1/\lambda_2, \ \lambda_3=1/\lambda_4$ so that if $\lambda_j(\veps)$ is real the energy $\veps_j(p)$ is in the gap. Otherwise, the eigenvalues consist of two pairs of conjugate complex numbers lying on the unit circle, re-numbered as $\lambda_1=1/\lambda_1^*, \lambda_2=1/\lambda_2^*$.  Consequently, there are {\it two symmetric dispersion curves} $\veps_1(p)=\veps_1(-p)$ and $\veps_2(p)=\veps_2(-p)$ corresponding to the two SO split levels. As we shall see below, for fixed $k_y=0$ and RSOC strength $\lambda \to 0$, the two curves coincide, forming valence and conduction bands that display a Dirac point at $p=0$, with linear dispersion $\veps_j(p)=\veps_j(0) +(-1)^j a |p|$ for small $p$ (where $a>0$ is a real constant), see Fig.~\ref{Fig1}(a).  The more intriguing case $\lambda>0$ is discussed below. We are now in a position to present our results based on the above formalism.   

{\it Choice of parameters}:
Our objectives are twofold: (1) to elucidate the Bloch dispersion and its dependence on the RSOC strength $\lambda$ (tunable by the electric field). (2) To calculate wave functions and spin-related observables, to asses their space dependence and their response to variation of $\lambda$. As we hope to enrich our understanding of graphene spintronics, it is important to choose potential parameters $u_0,d,\ell$ and RSOC strength $\lambda$ in accordance  with experimental capability. 

Below, the lengths $x,y,d,...$ are given in nm, and energies $\veps, u_0$ $\lambda$ as well as the wave numbers $k_x, k_y, q_x$ (introduced above) are given in (nm)$^{-1}$, [1 (nm)$^{-1}$ corresponds to 659.107 meV].  The size of $\lambda$ is dictated by experiments on Rashba spin-splitting in SLG. In Ref.~\cite{Gmitra}, it is shown that $\lambda$ is in the order of fraction of 1 meV. Here we let $0<\lambda \le$ 0.0016 nm$^{-1}$ ($0<\lambda \le 1.01544$ meV).  It is also required that the wave numbers $k_{xn}$ and $q_{xn}$ should be real [see Eq.~(\ref{3kn})]. For $k_y=0$, this implies $\veps>2 \lambda$ (for real $k_{xn}$) and  $(u_0-\veps)^2 + 2 \lambda (u_0-\veps)>0$ for $q_{xn}$. Finally, for simplicity, we consider forward scattering, $k_y=0$. The case $k_y \ne 0$ will be explored in a future communication. (Note that it is experimentally difficult to tune $k_y$ for fixed Fermi energy $\veps$). In summary: (1) The fixed parameters are: $k_y=0, u_0 = 98.85$ meV, $d=200$ nm, $\ell=260$ nm. (2) In the calculations of the spectrum the Bloch energy is varied in the interval $[2 \lambda, u_0/2]$. (3) In the calculations of the spin observables, the Fermi energy is fixed at 0.0243 (nm)$^{-1}=13.2$ meV. (4) Bloch spectrum and spin observables are calculated for $\lambda=0.00016$ (nm)$^{-1}=0.10544$ meV (nearly $\lambda = 0$),  and $\lambda=0.0016$ (nm)$^{-1}=1.0544$ meV.

{\it Results}:
In the series of figures below we show our results for the Bloch spectrum, the charge density,  and the non-diagonal elements of the spin current density. 
It is argued that the charge current density and the spin density vector vanish. 
Expressions for all these quantities are given below.

First, we discuss the Bloch spectrum. 
In Fig.~\ref{Fig1}(a) the dispersion curves $\veps(p)$ are shown for very small (actually vanishing) RSOC strength $\lambda=0.00016$ (nm)$^{-1}$=0.10054 meV. It consists of two (virtually) degenerate levels in the valence and the conduction bands with a single Dirac point at $p=0$. Strictly speaking, the periodic potential is 1D, so we should refer this linear dispersion as a Dirac triangle and not a Dirac cone. As we increase $\lambda$ to 0.0016(nm)$^{-1}=1.0054$ meV, the pattern is unusually modified as shown in Fig.~\ref{Fig1}(b). To explain what happens, it is useful to plot the inverse function $\sin p(\veps)R$ as function of $\veps$ (restricted to positive $p$ for simplicity). In Fig.~\ref{Fig1}(c) it is shown for $\lambda \to 0$, where the two $p$ levels coincide and form a Dirac point with linear dispersion. For $\lambda>0$ the red $p$ level ``pulls the Dirac point up'', and the two blue $p$ levels repel each other. As a result, (taking into account the symmetric pattern for $p<0$) it implies that RSOC causes level repulsion in both energy (except at the Dorac points) and momentum.  The single Dirac point at $p=0$ is now split into a couple of Dirac points $\pm p \ne 0$. But the dispersion at these two Dirac points remains linear, unlike in the pattern encountered in bulk SLG \cite{Jaros}. From the point of view of band structure, the central rhombus in Fig.~\ref{Fig1}(b) specifies a narrow ``semi-metallic band'' between the valence and conduction bands.

\begin{figure}[htb]
\centering
\subfigure[]
{\includegraphics[width=0.850\linewidth]{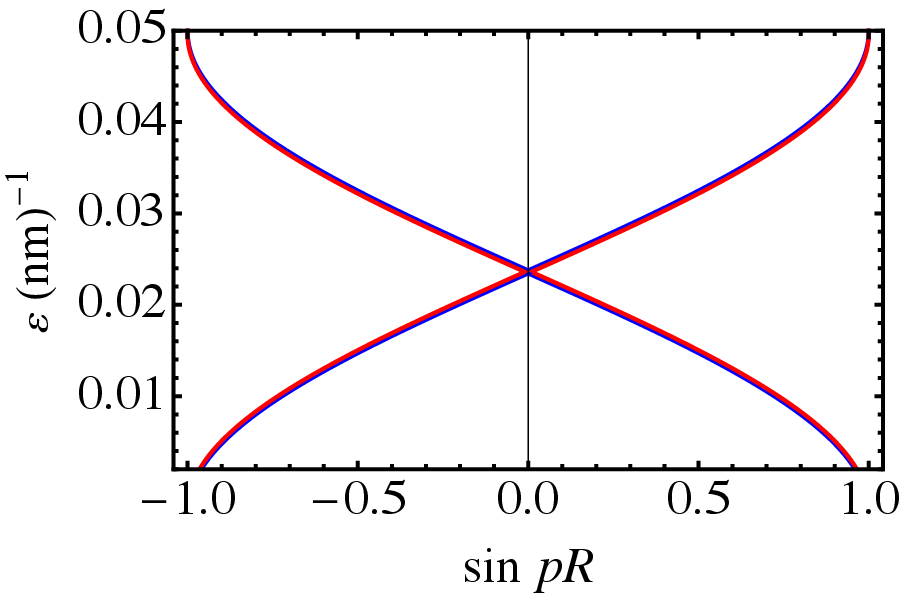}}
 \subfigure[]
{\includegraphics[width=0.850\linewidth]{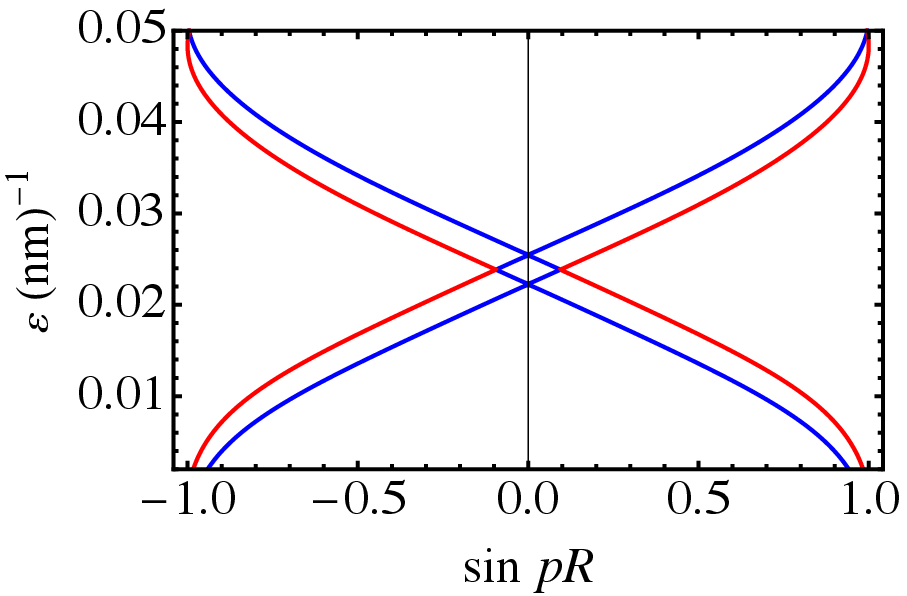}} 
 \subfigure[]
{\includegraphics[width=0.850\linewidth]{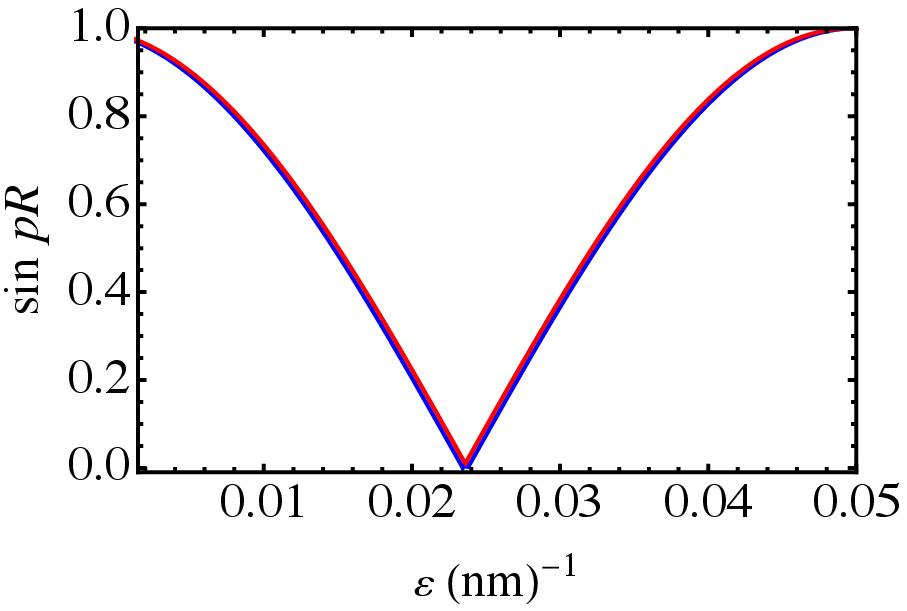}} 
 \subfigure[]
{\includegraphics[width=0.850\linewidth]{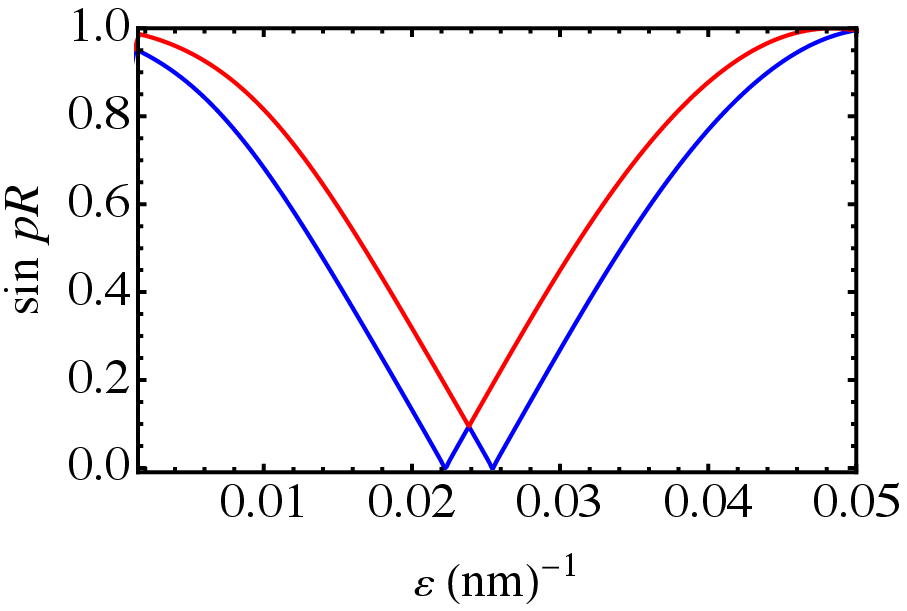}} 
\caption{\footnotesize (a) Bloch spectrum at $\lambda = 0.00016$(nm)$^{-1}=0.10054$ meV . (b) Bloch spectrum at $\lambda=0.0016$ (nm)$^{-1}=1.0054$ meV has {\it two Dirac points}. (c) and (d) Compare the inverse function $\sin p(\veps)R$: (c) is for $\lambda = 0.00016$(nm)$^{-1}=0.10054$ meV  and  (d) is for $\lambda=0.0016$ (nm)$^{-1}=1.0054$ meV.   These show that the effect of RSOC is to cause energy and momentum splitting of the two SO levels (prevailing at $\lambda=0$) while maintaining the Dirac points. 
}
\label{Fig1}
\vspace{-0.25in}
\end{figure}
Now we consider Bloch wave functions and derivation of local observables.
Calculations are carried out at a given energy $\veps$=0.025 nm$^{-1}$ that passes through the two Dirac points at $pR=\pm$0.13.  There are {\it four wave functions} $\{ \psi_{p_i}(x) \}, \ i=1,2,3,4$, corresponding to the four points $\{p_i \}$ at which the constant energy line crosses the four dispersion curves. The expressions of the wave functions are given in Eq.~(\ref{SO3}), wherein the coefficients $\{ a_{ns} \}$ are the component of the vector ${\bf a}$ [defined in Eq.~(\ref{compact-ab})] that is an eigenvector of $T$ with eigenvalue $e^{i p_i R}$. Similarly, the coefficients $\{ b_{ns} \}$ are the component of the vector ${\bf b}$ defined after Eq.~(\ref{SO7}).

An operator $\hat {O}$, is representable as 4$\times$4 hermitian matrix in ${\bm \sigma} \otimes {\bm \tau}$ (spin $\otimes$ isospin) space. Local observables are obtained by
\begin{equation} \label{Observable}
 O(x)=\frac{1}{4}\sum_{i=1}^4\psi_{p_i}^\dagger(x) \hat {O}\psi_{p_i}(x).
\end{equation}
(this is not an expectation value: observables may depend on $x$). Below we will consider operators of charge density, charge current (or velocity), spin density and spin current density, and check the space dependence of the corresponding observables.
 
For the charge density, the relevant operator is $\hat { I}_{4 \times 4}$ and the density is then $\rho(x) = \frac{1}{4}\sum_{i=1}^4\psi_{p_i}^\dagger(x)\psi_{p_i}(x)$. $\rho(x)$ is shown in Figs.~\ref{Fig2}(a) for $\lambda=0.00016$ (nm)$^{-1}$ and \ref{Fig2}(b) for $\lambda=0.0016$ (nm)$^{-1}$.  Note the  concentration of oscillations around  1. The reason is that  Bloch waves propagate in the  longitudinal direction (recall that $k_y=0$). In the absence of RSOC the Klein paradox implies that transmission through a barrier is unimpeded. As shown in Ref.~\cite{AB3}, in the presence of RSOC the transmission is still high but  not perfect. Increasing $\lambda$ implies larger oscillation amplitudes.  This is manifested here by noting that the amplitude of oscillations of the density at higher $\lambda$ as shown in Fig.~\ref{Fig2}(b), is larger than those for $\lambda \to 0$ in Fig.~\ref{Fig2}(a). The higher frequency in the barrier region $0<x<d$ (compared with the spacing region $-\ell < x <0$) reflects the inequality of wave numbers $q_{xn}>k_{kn}$, see Eq.~(\ref{3kn}).

Next we consider the velocity operator (which is also the charge current),
\begin{equation} \label{velocity}
 \hat{\bf V}={\bf I}_{2 \times 2} \otimes {\bm \tau}. 
 \end{equation}
As expected, we find that $V_x=0$, due to left-right symmetry. Also, $V_y=0$ because we have chosen $k_y=0$. However, 
the velocity operator will contribute to the spin current density (see below). 
\begin{figure}[htb]
\centering
\subfigure[]
{\includegraphics[width=0.80\linewidth]{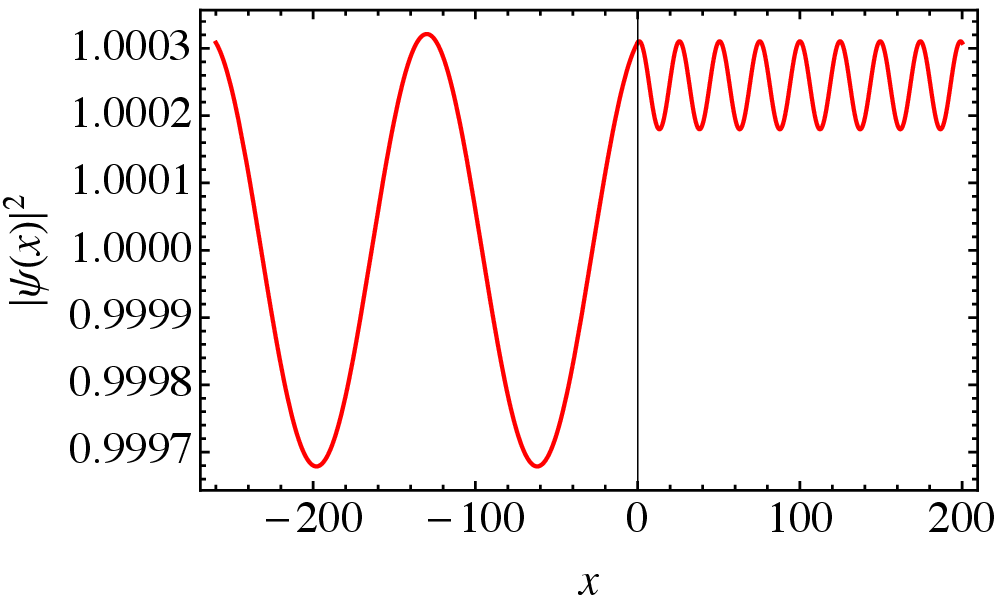}}
 \subfigure[]
{\includegraphics[width=0.8\linewidth]{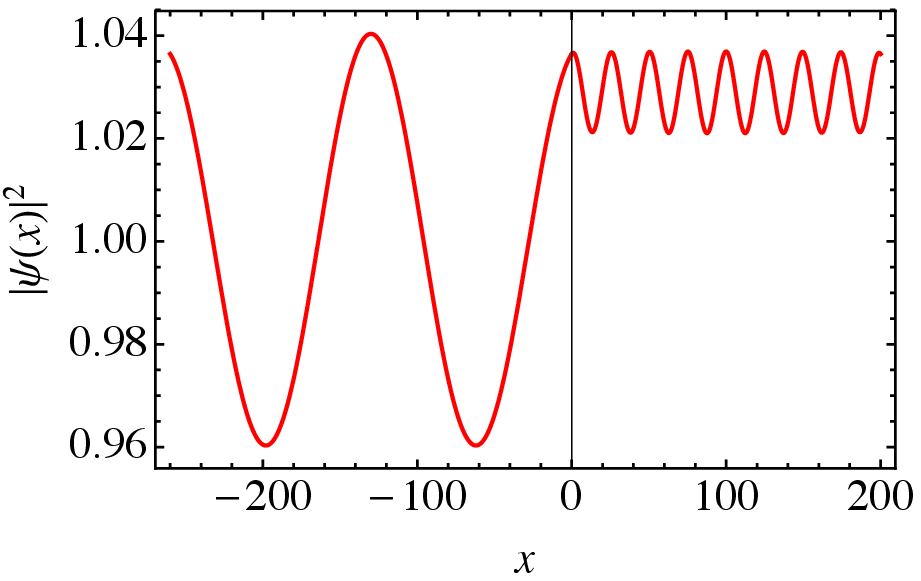}} 
\caption{\footnotesize Density $\rho(x)$ in the unit cell $-l < x <d$ for (a) $\lambda = 0.00016$ (nm)$^{-1} = 0.10054$ meV and (b) $\lambda=0.0016$ (nm)$^{-1} = 1.0054$ meV.
}
\label{Fig2}
\vspace{-0.2in}
\end{figure}

The spin density operators $\hat{\bf S}$ (from which the spin density observables ${\bf S}(x)$ are derived via Eq.~(\ref{Observable})), are given by, 
$\hat{\bf S}=(\hat{S}_x,\hat{S}_y,\hat{S}_z)
=\tfrac{1}{2} \hbar {\bm \sigma} \otimes {\bf I}_2.$
The unit of the observable ${\bf S}(x)$ is $S_0=\hbar/A$. But in the present case, it is found that ${\bf S}=0$. For $S_x$, it is expected that there is no spin density along the direction of motion. For $S_z$, it is expected that there is no spin density along the direction of motion outside the SLG plan. For $S_y$, there is cancellation between the four contributions in Eq.~(\ref{Observable}). 


Now let us focus on spin-current density. The corresponding operator is $\hat{\bf J}$ (a tensor) from which the observed components of the spin current density observables $J_{ij}(x)$ are  derived via Eq.~(\ref{Observable})), is defined as 
\begin{equation} \label{Jop}
 \hat{J}=\frac{1}{2} \{ \hat{\bf S},\hat{\bf V} \}, 
\end{equation} 
where $\hat{\bf S}$ is the spin density operator defined above, and $\hat{\bf V}= {\bf I}_2 \otimes {\bm \tau}$ is the velocity operator defined in Eq. (\ref{velocity}).  In Eq.~(\ref{Jop}), $i=1,2,3 = x,y,z$ specifies the polarization direction, and $j=1,2=x,y$ specifies the axis along which electrons propagate.  The unit of spin current density observables $J_{ij}$ is $J_0=S_0 v_F=\gamma /A$ meV/nm.
 
Our calculations show that the non-zero spin current density observables are the non-diagonal elements of the spin current density observable, explicitly, $J_{xy}(x)=J_{yx}(x), J_{zx}(x)$ and $J_{zy}(x)$. They are shown in Figs.~\ref{Fig3}, \ref{Fig4} and \ref{Fig5} respectively for $\lambda$=0.00016(nm)$^{-1}$=0.10054 meV in panel (a) and $\lambda=0.0016$ (nm)$^{-1}$ = 1.0054 meV in (b). Note that (1) Despite the fact that ${\bf V}={\bf S}=0$, the spin current density does not vanish. (2) Increasing $\lambda$ by a factor 10  increases the amplitudes of the spin current density by a factor of about 15 for $J_{xy}$ and $J_{zx}$ and about 50 for $J_{zy}$. (3) The spin current densities have a rich space dependence implying a non-zero torque, see below. 

\begin{figure}[htb]
\centering
\subfigure[]
{\includegraphics[width=0.80\linewidth]{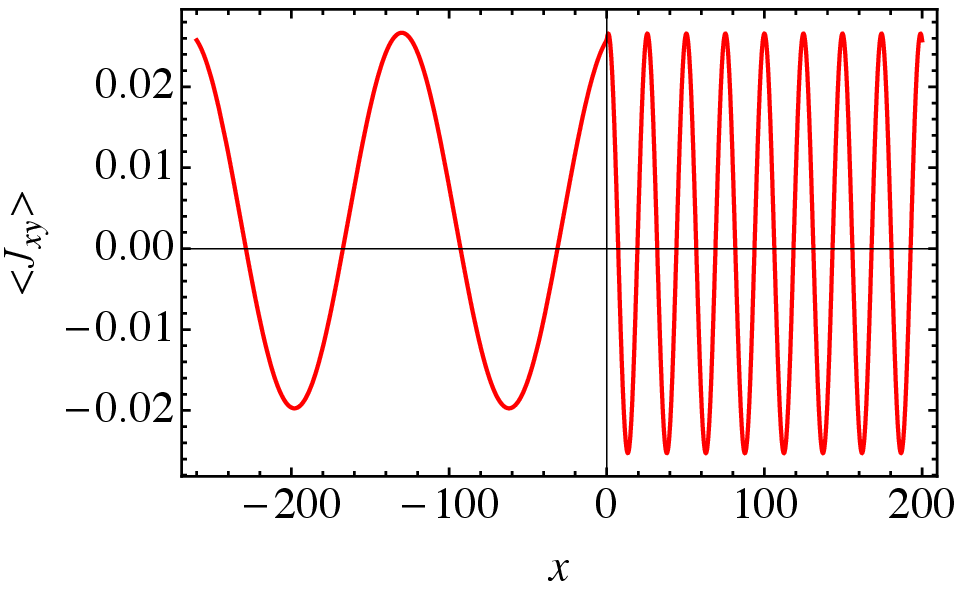}}
 \subfigure[]
{\includegraphics[width=0.8\linewidth]{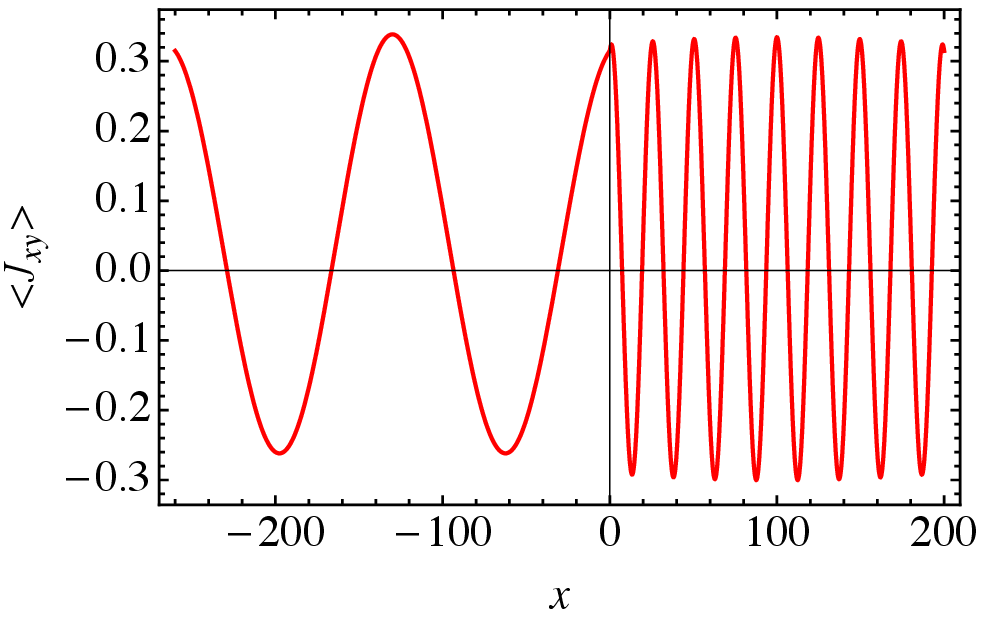}} 
\caption{\footnotesize Spin current density $J_{xy}(x)$ in the unit cell $-l < x <d$ for
(a)  $\lambda = 0.00016$ (nm)$^{-1}$(nm)$^{-1}=0.10054$ meV and (b) $\lambda=0.0016$ (nm)$^{-1}=1.0054$ meV. The ratio of amplitude $J_{xy}$ in (b) to $J_{xy}$ in (a) is abut 15. Note that $J_{yx}(x)=J_{xy}(x)$. 
}
\label{Fig3}
\end{figure}

\begin{figure}[htb]
\centering
\subfigure[]
{\includegraphics[width=0.80\linewidth]{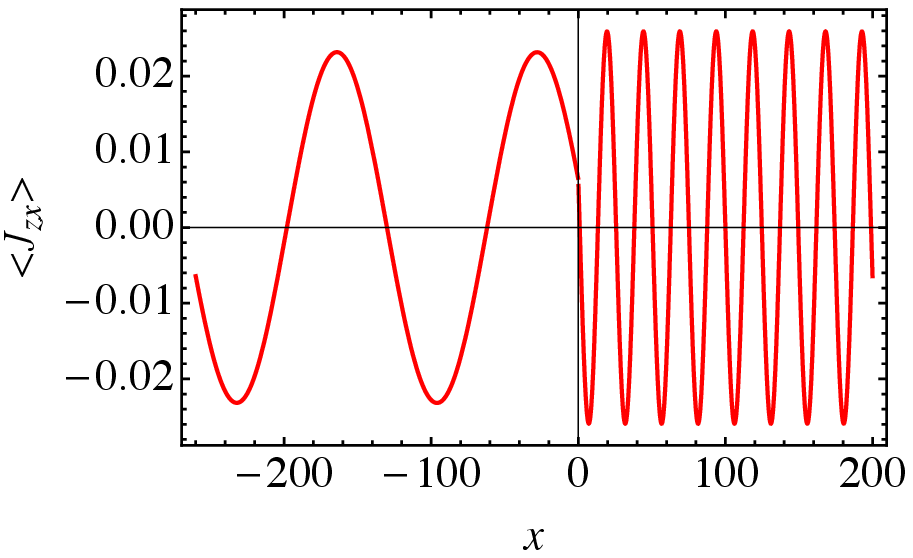}}
 \subfigure[]
{\includegraphics[width=0.8\linewidth]{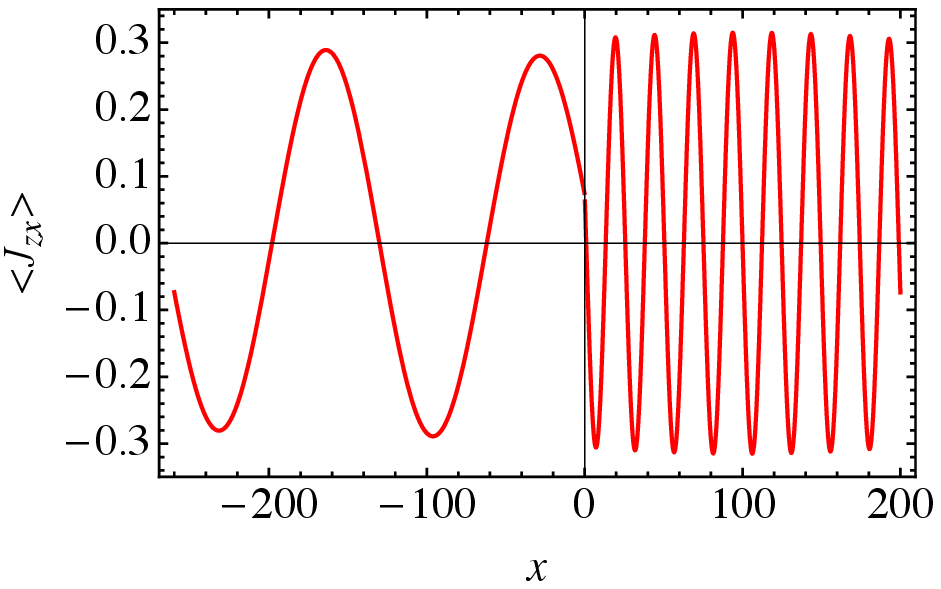}} 
\caption{\footnotesize Spin current density $J_{zx}(x)$ in the unit cell $-l < x <d$ for 
(a)  $\lambda = 0.00016$ (nm)$^{-1}$(nm)$^{-1}=0.10054$ meV and (b) $\lambda=0.0016$ (nm)$^{-1}=1.0054$ meV. The ratio of amplitude $J_{zx}$ in (b) to $J_{zx}$ in (a) is about 15.
}
\vspace{-0.25in}
\label{Fig4}
\end{figure}

\begin{figure}[htb]
\centering
\subfigure[]
{\includegraphics[width=0.80\linewidth]{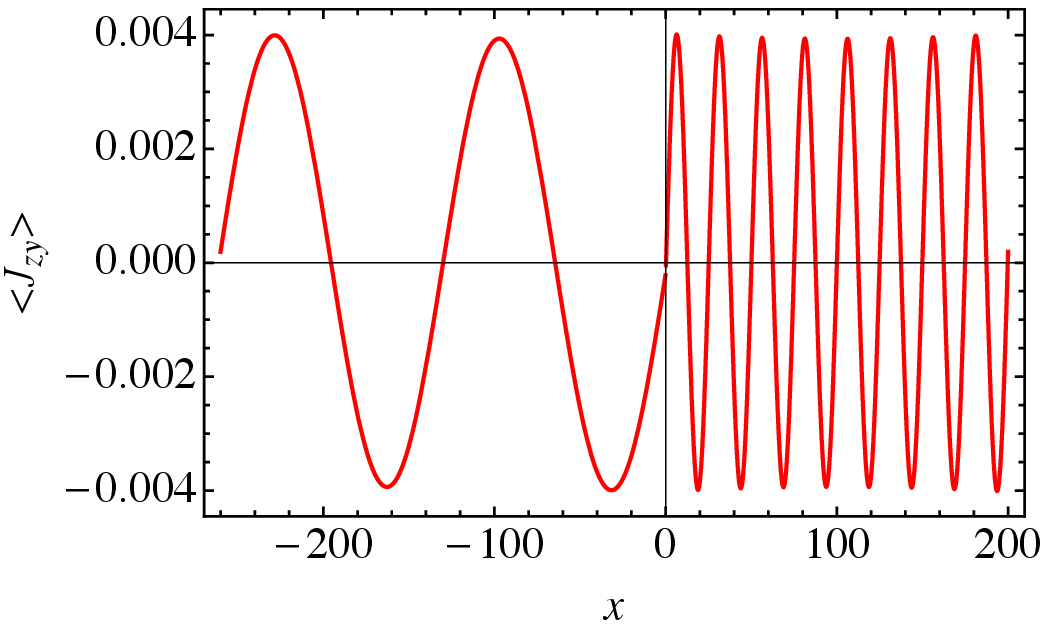}}
 \subfigure[]
{\includegraphics[width=0.8\linewidth]{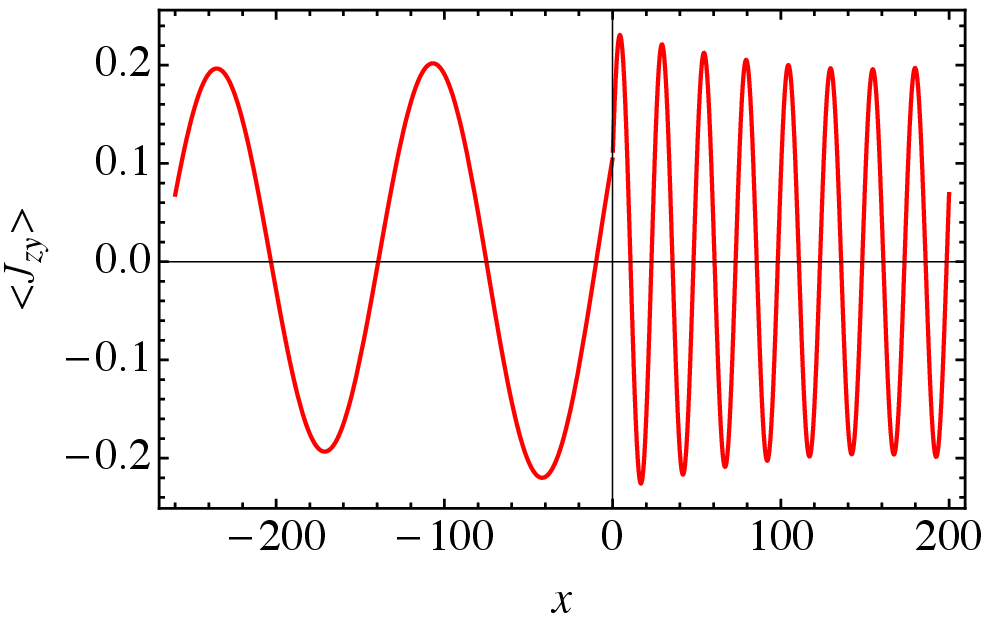}} 
\caption{\footnotesize Spin current density $J_{zy}(x)$ in the unit cell $-l < x <d$ for (a)  $\lambda = 0.00016$ (nm)$^{-1}$(nm)$^{-1}=0.10054$ meV and (b) $\lambda=0.0016$ (nm)$^{-1}=1.0054$ meV.  The ratio of amplitude $J_{zy}$ in (b) to $J_{zy}$ in (a) is about 50.
}
\label{Fig5}
\vspace{-0.25in}
\end{figure}

The spin current density was calculated in bulk SLG in Ref.~\cite{Zhang}. The authors found that (1) $J_{xx}=J_{yy}=J_{zx}=J_{zy}=0$, (2) $J_{xy}=-J_{yx}$, and (3) the spin currents are not space dependent (see Eq.~(5) in Ref.~\cite {Zhang}).  In our calculations it is shown that in the presence of a 1D potential (where there is no rotational symmetry around the $z$-axis), the symmetry relation  (valid in bulk SLG \cite{Zhang}) is reversed, $J_{xy}=+J_{yx}$. Moreover, although the value of $\lambda$ used in our calculations is about two orders of magnitude smaller than that used in Ref.~\cite{Zhang}, the size of the spin current densities in both systems {\color{blue} are} the same order of magnitude. Another noticeable difference from SLG is that in the present system, spin current densities are space dependent and the divergence of the spin current density does not vanish. The continuity equation for the spin current density must contain a spin torque density term \cite{Niu}.  As we have shown in Ref.~\cite{AB3}, for spin current density that depends only on $x$, the component $J_{ix}(x)$ have  non-zero torque.  In the present case these are $J_{yx}$ and $J_{zx}$. 

{\it Summary and Conclusion}: 
The  Klein  paradox in SLG occurs when an electron at the Fermi energy $\veps$ tunnels through a 1D potential barrier of height $u_0$ (which can be experimentally controlled by a gate voltage) in the region $u_0>\veps>0$. When, in addition, a uniform perpendicular electric field ${\bf E} = E_0 \hat {\bf z}$ is applied, the role of electron spin enters due to RSOC. This system was  studied in relation to transmission \cite{Richter,Shakouri,AB2} and spin current densities \cite{AB2} with the quest to reveal interesting facets of graphene spintronics within a time-reversal invariant formalism. Its study is appealing due to the fact that $u_0$ and the RSOC strength $\lambda$ can be experimentally controlled, making it verifiable.

The present work targets graphene spintronics not through the properties of transmission, but rather through the properties of the stationary states.  It combines the four pillars of 2D Dirac electrons, Klein Paradox, Bloch theorem and RSOC, and establishes a theoretical framework with predictive power. It presents a plethora of observables that can be experimentally tested.  It is shown that RSOC results in an unusual Bloch dispersion band structure with two Dirac points and a narrow semi-metallic band between the valence and conduction bands. Spin observables are calculated and shown to have different properties than those found in bulk SLG. In particular, spin current density exists also if the polarization direction is perpendicular to the graphene plane. Moreover, despite the upper (experimental) limit on the strength $\lambda$ of the RSOC, the size of the spin current density is not small. In addition, the spin current density has non-trivial space dependence along the periodic lattice direction, implying the occurrence of finite and oscillating spin torque density \cite{Niu}.

This work is partially motivated by the quest for constructing spintronic devices without the use of an external magnetic field or magnetic materials (in addition to the many references mentioned above, see also Refs.~\cite{Hatano, Matityahu, AB1}).  We hope that our results advance this goal. The sensitivity of the spectrum $\veps(p)$ and the components $J_{ij}(x)$  of the spin current density tensor to variation of the RSOC strength $\lambda$ are particularly promising aspects in this regard.

{\it Acknowledgement}: Discussions with J. Nitta, J. Fabian, K. Richter and M. H. Liu are highly appreciated, and were indispensable for understanding some crucial issues. 


\end{document}